# Chapter 9
# Cryogenics for the HL-LHC

*S. Claudet[*] and L. Tavian*

CERN, Accelerator & Technology Sector, Geneva, Switzerland

## 9    Cryogenics for the HL-LHC

### 9.1    Introduction

The upgrade of the cryogenics for the HL-LHC will consist of the following:

- The design and installation of two new 1.9 K cryogenic plants at P1 and P5 for high luminosity insertions. This upgrade will be based on a new sectorization scheme aimed at separating the cooling of the magnets in these insertion regions from the arc magnets, and on a new cryogenic architecture based on electrical feedboxes located at ground level and vertical superconducting links.

- The design and installation of a new 4.2 K cryogenic plant at P4 for the Superconducting Radio Frequency (SRF) cryomodules and other future possible cryogenic equipment (e-lens, RF harmonic system).

- The design of new cryogenic circuits at P7 for the HTS links and displaced current feedboxes.

- Cryogenic design support for the 11 T dipoles.

    Figure 9-1 shows the overall LHC cryogenic layout, including the upgraded infrastructure.

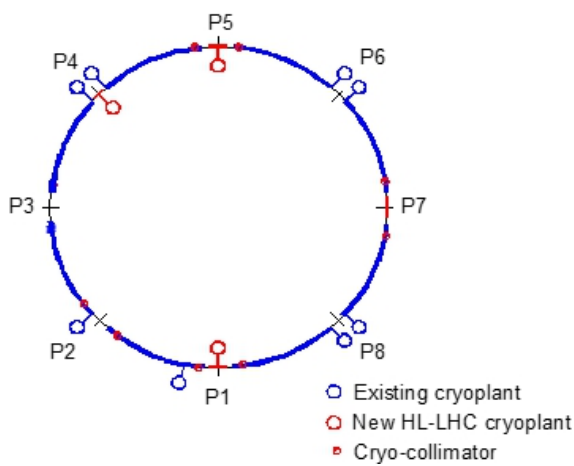

Figure 9-1: Overall LHC cryogenic layout, including the upgraded infrastructure

---

[*] Corresponding author: Serge.Claudet@cern.ch

## 9.2 LHC machine upgrades

9.2.1 Upgraded beam parameters and constraints

The main parameters impacting on the cryogenic system are given in Table 9-1. With respect to the nominal beam parameters, the beam bunch population will double and the luminosity in the detectors of the high luminosity insertions at P1 and P5 will be multiplied by a factor 5.

Table 9-1: LHC upgraded beam parameters for 25ns bunch spacing

| Parameter | Units | Nominal | Upgrade |
|---|---|---|---|
| Beam energy, $E$ | [TeV] | 7 | 7 |
| Bunch population, $N_b$ | [protons/bunch] | $1.15 \times 10^{11}$ | $2.2 \times 10^{11}$ |
| Number of bunches per beam, $n_b$ | - | 2808 | 2748 |
| Luminosity, $L$ | [cm$^{-2}$ s$^{-1}$] | $1 \times 10^{34}$ | $5 \times 10^{34}$ |
| Bunch length | [ns] | 1.04 | 1.04 |

These upgraded beam parameters will introduce new constraints to the cryogenic system.

- The collimation scheme must be upgraded by adding collimators to the continuous cryostat close to P2 and P7, and possibly also P1 and P5. The corresponding integration space must be created by developing shorter but stronger 11 T dipoles. As the new collimators will work at room temperature, cryogenic bypasses are required to guarantee the continuity of the cryogenic and electrical distribution. Figure 9-2 shows the nominal and upgraded layouts of the continuous cryostat. Halo control for the HL-LHC may require the installation of hollow electron lenses at P4, making use of a superconducting solenoid. While not yet in the HL-LHC baseline, this device may be the best option for controlling particle diffusion by depopulating the halo of the high-power hadron beams, thereby avoiding uncontrolled losses during critical operations such as the squeeze. Figure 9-3 shows the nominal and upgraded layouts of the P4 insertion region (IR4), anticipating the installation of an e-lens and a new SRF system.

- The increase of the level of radiation to the electronics could possibly require relocating power convertors and related current feedboxes to an access gallery at P7 and at ground level at P1 and P5. New superconducting links will be required to connect the displaced current feedboxes to the magnets. Figure 9-4 and Figure 9-5 show the nominal and upgraded layouts of IR1, IR5, and IR7.

- To better control the bunch longitudinal profile, reduce heating and improve the pile-up density, new cryomodules of 800 MHz RF cavities could be added to the existing 400 MHz cryomodules at P4 creating a high-harmonic RF system (see Figure 9-3). Actually, discussions are underway to see if a better scheme would be the installation of a new 200 MHz SRF system, rather than the 800 MHz. From the cryogenic point of view the requests are similar, so we will consider below the 800 MHz system that is in an advanced phase of study.

- To improve the luminosity performance by addressing the geometric luminosity reduction factor and possibly allowing the levelling of the luminosity, cryomodules of crab cavities (CC) will be added at P1 and P5 (see Figure 9-5).

- Finally, the matching and final focusing of the beams will require completely new insertion assemblies at P1 and P5 (see Figure 9-5).

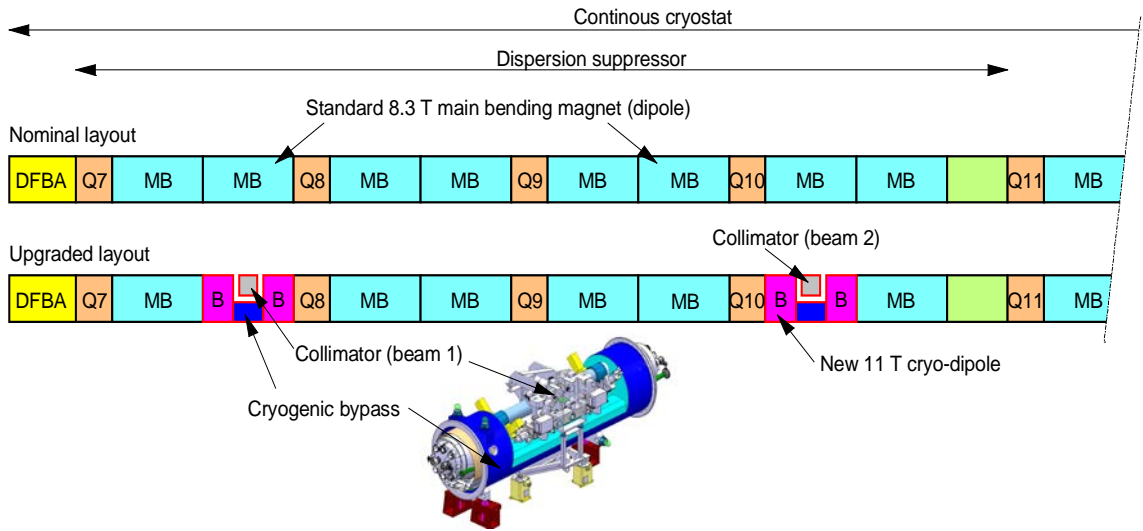

Figure 9-2: Upgraded layout of the continuous cryostat at P2 (as well at P1, P5, and P7)

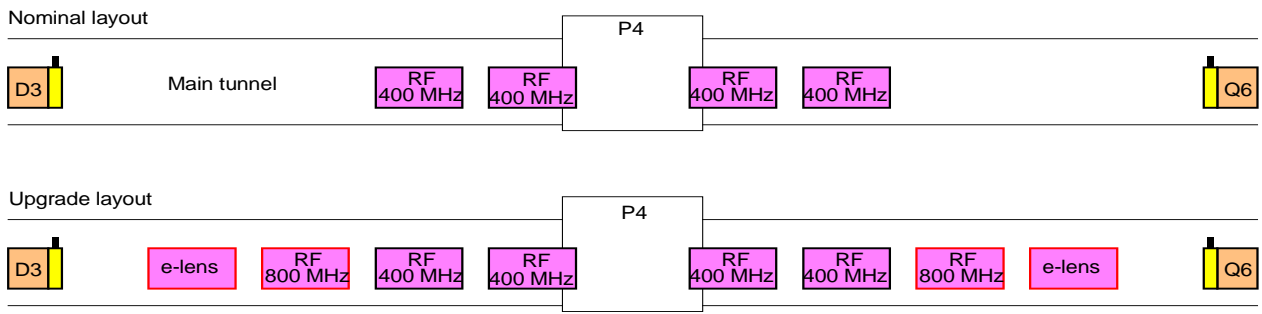

Figure 9-3: Possible upgraded layout of the P4 insertion region

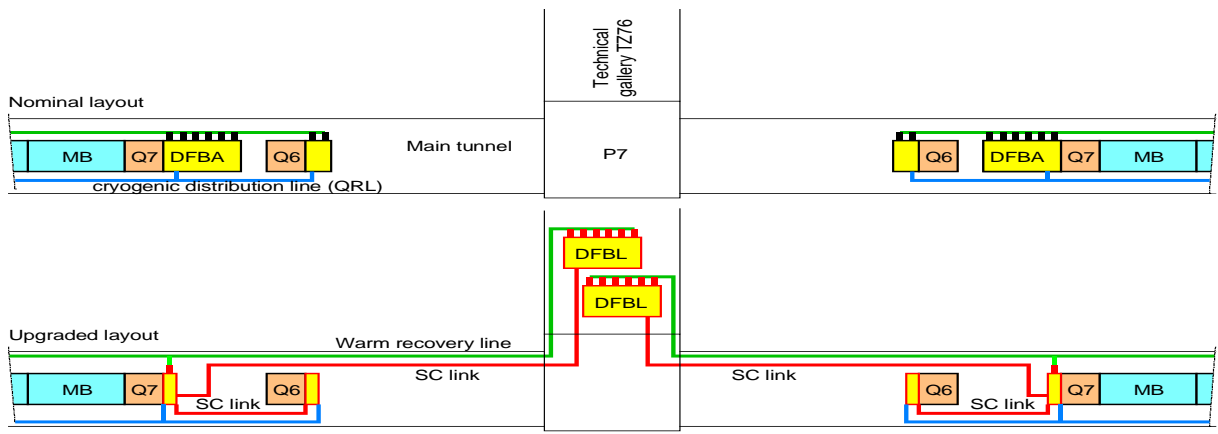

Figure 9-4: Upgraded layout of the P7 insertion region

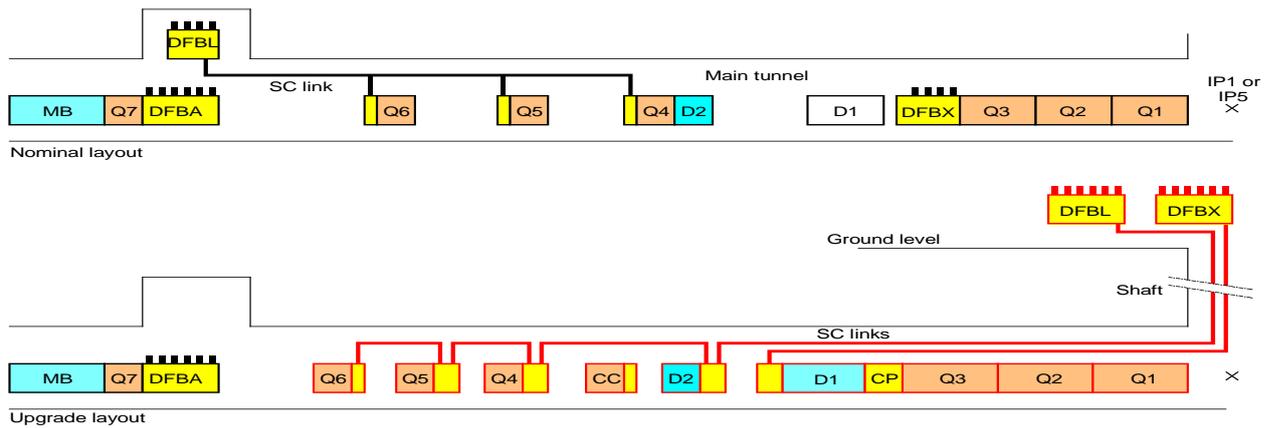

Figure 9-5: Upgraded layout of the P1/P5 insertion region (half insertion)

## 9.3 Temperature level and heat loads

Heat loads to the cryogenic system have various origins and uncertainties. As done for the LHC and clearly described in the design report, two categories of heat loads are considered: static heat loads ($Q_{static}$) to be compensated just to reach the desired temperature level, and dynamic heat loads ($Q_{dynamic}$) due to energising or circulating beams. These heat loads are primarily considered without contingency to avoid piling-up margins. However, the cooling capacity to be installed has to include margins that vary for the static and dynamic heat loads to properly allow the nominal beam scenario. This margin vanishes for the ultimate beam scenario.

In Table 9-2 the static heat in-leaks are reported, for different temperature levels. For new equipment, the thermal performance of supporting systems, radiative insulation and thermal shields is considered identical to that of existing LHC equipment.

Table 9-2: Static heat in-leaks of HL-LHC machine (without contingency)

| Temperature [k] | Equipment | Unit | Nominal | Upgrade |
|---|---|---|---|---|
| 4.6–20 | Beam screen circuit (arc + DS) | [mW/m] | 140 | 140 |
|  | Beam screen circuit (IT) | [mW/m] | 125 | 125 |
|  | Beam screen circuit (MS) | [mW/m] | 578 | 578 |
| 1.9 | Cold mass (arc + DS) | [mW/m] | 170 | 170 |
|  | Cold mass (IT) | [mW/m] | 1250 | 1250 |
|  | Crab cavities | [W per module] | 0 | 25 |
| 4.5 | Cold mass (MS) | [mW/m] | 3556 | 3556 |
|  | 400 MHz RF module | [W per module] | 200 | 200 |
|  | 800 MHz RF module | [W per module] | 0 | 120 |
|  | Electron-lens | [W per module] | 0 | 12 |
| 20–300 | Current lead | [g/s per kA] | 0.035 | 0.035 |

Table 9-3 gives the dynamic heat loads expected for the HL-LHC. The main concern is electron-cloud impingement on the beam screens, which can only be reduced by efficient beam scrubbing (dipole off) of the beam screens. This remains to be demonstrated. Without efficient beam scrubbing (dipole on), the e-cloud activity will remain high (more than 4 W/m and per beam) in the arcs and dispersion suppressors (DS). This heat deposition corresponds to about twice the local cooling limitation given by the hydraulic impedance of the beam screen cooling circuits. In addition, the corresponding integrated power over a sector (more than 25 kW) is not compatible with the installed capacity of the sector cryogenic plants. For e-cloud deposition in the arcs and dispersion suppressors, efficient (dipole off) or inefficient (dipole on) beam scrubbing is considered.

Table 9-3: Dynamic heat loads on HL-LHC machine (without contingency)

| Temperature [k] | Equipment | Unit | Nominal | Upgrade |
|---|---|---|---|---|
| 4.6–20 | Synchrotron radiation (arc + DS) | [mW/m per beam] | 165 | 310 |
| | Image current (arc + DS + MS) | [mW/m per beam] | 145 | 522 |
| | Image current (IT low-luminosity) | [mW/m] | 475 | 1698 |
| | Image current (IT high luminosity) | [mW/m] | 166 | 596 |
| | E-clouds (arc + DS) (dipole off) | [mW/m per beam] | 271 | 41 |
| | E-clouds (arc + DS) (dipole on) | [mW/m per beam] | 4264 | 4097 |
| | E-clouds (IT high luminosity) | [mW/m] | 5500 | 9455 |
| | E-clouds (IT low-luminosity) | [mW/m] | 5500 | 5500 |
| | E-clouds (MS) | [mW/m per beam] | 2550 | 383 |
| | Secondaries (IT beam screen P1 andP5) | [W per IT] | 0 | 650 |
| 1.9 | Beam gas scattering | [mW/m per beam] | 24 | 45 |
| | Resistive heating in splices | [mW/m] | 56 | 56 |
| | Secondaries (IT cold mass P1 and P5) | [W per IT] | 155 | 630 |
| | Secondaries (DS cold mass P1 and P5) | [W per DS] | 37 | 185 |
| | SCRF crab cavities | [W per module] | 0 | 24 |
| 4.5 | SCRF 400 MHz | [W per module] | 101 | 366 |
| | SCRF 800 MHz | [W per module] | 0 | 183 |
| | E-lens | [W per module] | 0 | 2 |
| 20–300 | Current lead | [g/s per kA] | 0.035 | 0.035 |

The beam screens of the new inner triplets at P1 and P5 will be protected by tungsten shielding that will be able to absorb about half of the energy deposited by collision debris escaping the high luminosity detectors. For simplicity at this stage, beam screen loads were considered to be between 4.6 K and 20 K as for the current LHC. However, the large dynamic power to be extracted could force consideration of the next possible temperature range compatible with beam vacuum requirements, i.e. the range 40 K to 60 K. Despite this thick W-shielding, the 1.9 K load, i.e. the energy that collision debris deposited onto the magnet coil and cold mass, increases by four times with respect to the nominal LHC case. The W-shielding, in any case, reduces the overall refrigeration cost and increases the lifetime of the inner-triplet coils.

## 9.4 Impact on existing sector cryogenic plants

With new cryogenic plants dedicated to the cooling of cryogenic equipment in P1, P4, and P5, the cooling duty of the existing sector cryogenic plants will be reduced and more equally distributed. Figure 9-6 shows the required cooling capacities for the different temperature levels and compares them to the nominal cooling requirements and to the installed capacities. The low-load sectors equipped with upgraded ex-LEP cryogenic plants have lower installed capacity than the four cryogenic plants specially ordered for the LHC's high-load sectors. For the HL-LHC, sufficient capacity margin still exists providing that the beam scrubbing of dipole beam-screens is efficient (dipole off).

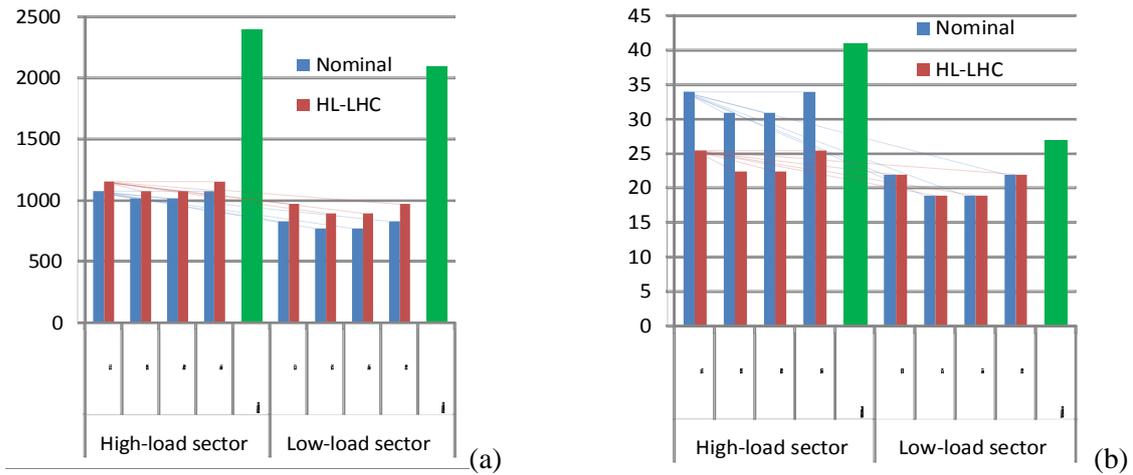
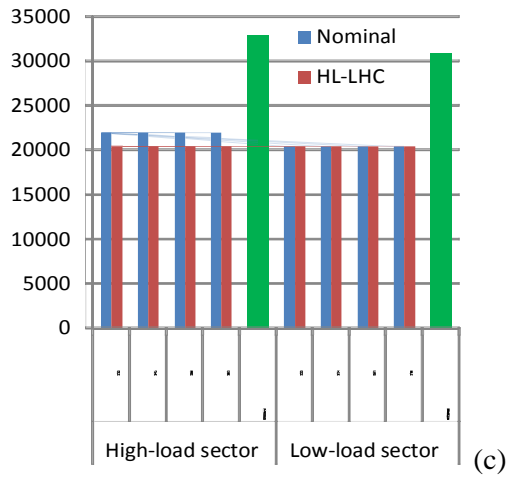
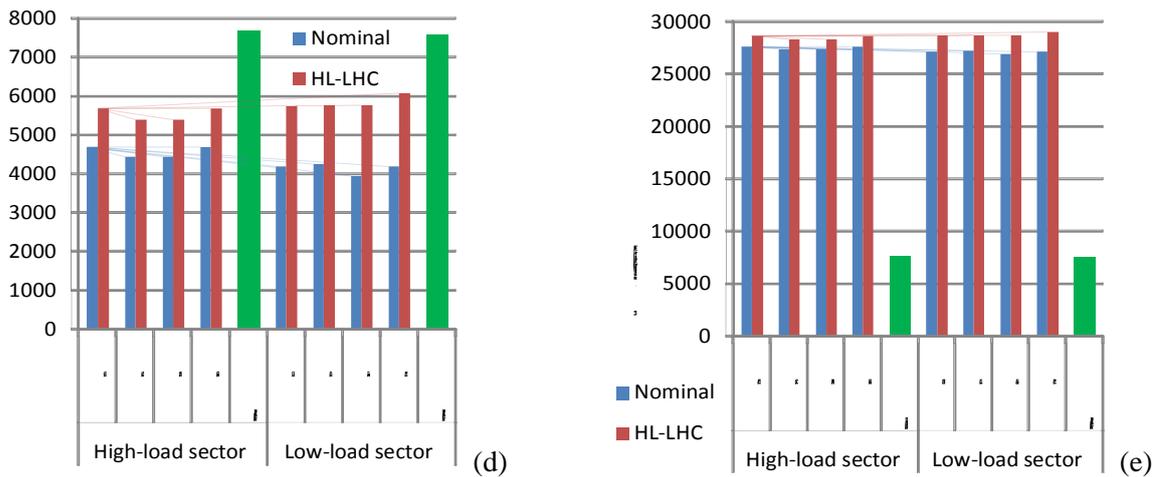

Figure 9-6: Cooling capacity requirement of sector cryogenic plants. (a) Cold mass; (b) current leads; (c) thermal shields; (d) beam screen (dipole off); (e) beam screen (dipole on).

## 9.5 New cryogenics for Point 4 insertion

Figure 9-7 shows the cryogenic architecture of the upgraded P4 insertion consisting of:

- a warm compressor station (WCS) located in a noise-insulated surface building and connected to a helium buffer storage;
- a lower cold box (LCB) located in the UX45 cavern and connected to a cryogenic distribution valve box (DVB), also located in the UX45 cavern;
- main cryogenic distribution lines connecting the cryomodules to the distribution valve box;
- auxiliary cryogenic distribution lines interconnecting the new infrastructure with the existing QRL service modules (SM) and allowing redundancy cooling with adjacent-sector cryogenic plants;
- a warm-helium recovery line network.

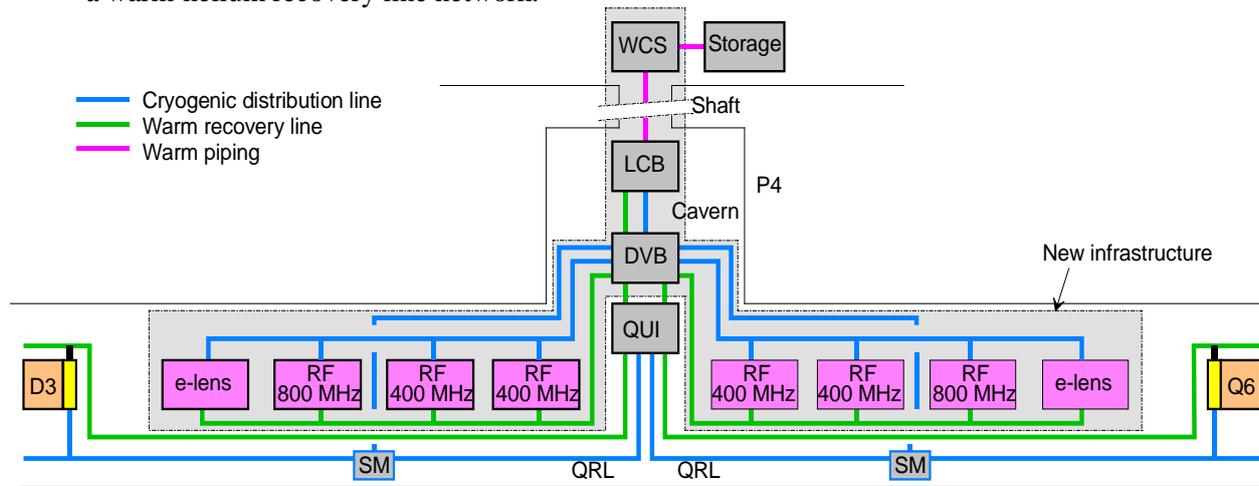

Figure 9-7: Upgraded cryogenic architecture at P4

Concerning the planned installed capacity ($Q_{installed}$) of the new cryogenic plant, some uncertainty ($f_u$) and overcapacity ($f_o$) margins have to be introduced as shown in Eq. 9.1:

$$Q_{installed} = f_o \times (Q_{static} \times f_u + Q_{dynamic}), \tag{9-1}$$

where $f_o = 1.5$ and $f_u = 1.5$.

Table 9-4 gives the installed capacity of the P4 cryogenic plant required at different temperature levels. The P4 cryogenic plant will require an equivalent capacity of about 6 kW at 4.5 K.

This is considered as the present baseline, with the evaluation of an alternative scenario for the refrigeration part. The alternative scenario would consist of an upgrade of one of the existing refrigerator of P4 to fulfil the required cooling capacity of existing SRF modules with sufficient margin, while keeping the baseline new distribution scenario. This modular and staged approach would allow the installation at a later stage of a new and dedicated refrigerator adapted to the loads presently under definition.

Table 9-4: Installed capacity requirements of the new cryogenic plant at P4

| Temperature level [K] | Static [W] | Dynamic [W] | Installed [W] | Equivalent installed capacity at 4.5 K [kW] | |
|---|---|---|---|---|---|
| 4.5 | 1144 | 1736 | 5223 | 5.6 | 5.8 |
| 50–75 | 1000 | 0 | 2250 | 0.2 | 5.8 |

## 9.6 New cryogenics for high luminosity insertions at Point 1 and Point 5

Figure 9-8 shows the proposed cryogenic architecture of the P1 and P5 high luminosity insertions consisting of:

- a warm compressor station (WCS) located in a noise-insulated surface building and connected to a helium buffer storage;
- an upper cold box (UCB) located in a ground-level building;
- a quench buffer (QV) located at ground level;
- one or two cold compressor boxes (CCB) in an underground cavern;
- two main cryogenic distribution lines (one per half-insertion);
- two interconnection valve boxes with existing QRL cryogenic line allowing redundancy with the cryogenic plants of adjacent sectors.

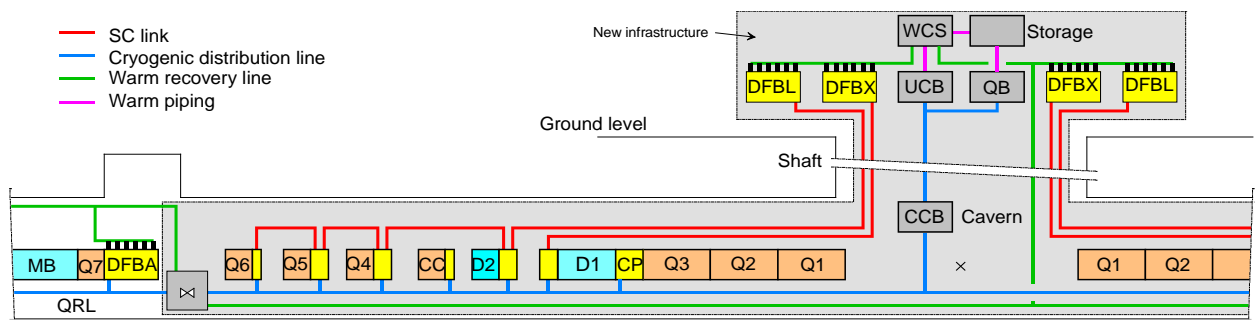

Figure 9-8: Upgraded cryogenic architecture at P1 and P5

Table 9-5 gives the installed capacity of the proposed P1 and P5 cryogenic plants required at the different temperature levels and using the same uncertainty and overcapacity margins as those used for P4. The cryogenic plants will require an equivalent capacity of about 18 kW at 4.5 K, including 3 kW at 1.8 K.

Table 9-5: Installed capacity requirements of the new cryogenic plants at P1 and P5

| Temperature level [K] | Units | Static | Dynamic | Installed | Equivalent installed capacity at 4.5 K [kW] | |
|---|---|---|---|---|---|---|
| 1.9 | [W] | 433 | 1380 | 3045 | 12 | 18 |
| 4.5 | [W] | 196 | 8 | 452 | 0.5 | 18 |
| 4.6–20 | [W] | 154 | 2668 | 4348 | 2.4 | 18 |
| 50–75 | [W] | 4900 | 0 | 7350 | 0.5 | 18 |
| 20–300 | [g/s] | 16 | 16 | 59 | 2.6 | 18 |

At P1 and P5 the superconducting magnets of the ATLAS and CMS detectors are cooled by dedicated cryogenic plants. A possible redundancy with detector cryogenic plants could be interesting in the event of a major breakdown of the detector cryogenic plants. The corresponding power requirements are about 1.5 kW at 4.5 K for CMS and 3 kW at 4.5 K for ATLAS.

The cooling capacity of 3 kW at 1.8 K is higher than the 2.4 kW installed capacity of an LHC sector, which corresponds to the present state-of-the-art for the cold compressor size. Consequently:

- larger cold compressors have to be studied and developed;
- or parallel cold compressor trains have to be implemented (one 1.5 kW train per half insertion);
- or duplication of the first stage of cold compression to keep the machine within the available size.

## 9.7 Building and general service requirement

Table 9-6 gives the buildings and general services at P1, P4, and P5 required by the cryogenic infrastructure. At P4, the required surface and volume for the warm compression station and for the cold box are respectively available in the existing SUH4 building and in the UX45 cavern.

Table 9-6: Building and general service requirement

| Cryogenic system | | | P1 and P5 | P4 |
|---|---|---|---|---|
| Warm compressor building | Surface | [m$^2$] | 700 | 400 |
| | Crane | [t] | 20 | 20 |
| | Electrical power | [MW] | 4.6 | 2.0 |
| | Cooling water | [m$^3$/h] | 540 | 227 |
| | Compressed air | [Nm$^3$/h] | 30 | 20 |
| | Ventilation | [kW] | 250 | 100 |
| | Type | - | Noise-insulated (~108 dB_A) | |
| Surface SD building | Surface | [m$^2$] | 30 × 10 | N/A |
| | Height | [m] | 12 | N/A |
| | Crane | [t] | 5 | N/A |
| | Electrical power | [kW] | 50 | N/A |
| | Cooling water | [m$^3$/h] | 15 | N/A |
| | Compressed air | [Nm$^3$/h] | 90 | N/A |
| Cavern | Volume | [m$^3$] | 200 | 300 |
| | Local handling | [t] | 2 | 2 |
| | Electrical power | [kW] | 100 | 20 |
| | Cooling water | [m$^3$/h] | 20 | 20 |
| | Compressed air | [Nm$^3$/h] | 40 | 30 |

## 9.8 Conclusions

The HL-LHC project will require a major cryogenic upgrade. The main challenges are given below.

- Cooling circuits for large heat deposition.
  - Up to 13 W/m on 1.9 K cold masses for heat extraction from SC cables and sufficient quench energy margin. Accurate heat flow calculation in coil and yoke cross-section must be developed.
  - Up to 23 W/m on inner-triplet beam screens possibly with a different operating range (40–60 K) and with a large dynamic range that will require specific cryogenic plant adaptation studies.
- Cooling of HTS SC links and current feedboxes.
- Cooling and pressure relief of crab cavities.
- Validation tests on SC link, crab cavities, magnets, beam screens, etc.
- Reactivation of the Heat Load Working Group.
- Quench containment and recovery.
- Larger 1.8 K refrigeration capacities beyond the present state-of-the-art.
- Large capacity (1500 W/3000 W) sub-cooling heat exchangers.
- Larger turndown capacity factor (up to 10) on the 1.8 K refrigeration cycle.